\documentclass[aps,prl,twocolumn,groupedaddress,showpacs]{revtex4}
\usepackage{graphicx}	% Include figure files
\usepackage{dcolumn}	% Align table columns on decimal point
\usepackage{bm}		% bold math
\usepackage{color}
\usepackage{ulem}

\begin{document}

\include{epsf}

%\preprint{}

\title{Giant formation rates of ultracold molecules via 
Feshbach Optimized Photoassociation} 
%\title{Feshbach Optimized PhotoAssociative Spectroscopy}



\author{Philippe Pellegrini, Marko Gacesa, and Robin C\^ot\'e}

%\thanks{}
\affiliation{Department of Physics, U-3046, University of Connecticut, 
             Storrs, CT, 06269-3046}

%\thanks{}


\date{\today}

\begin{abstract}

Ultracold molecules offer a broad variety of applications, 
ranging from metrology to quantum computing.
However, forming ``real" ultracold molecules, {\it i.e.}
in deeply bound levels, is a very difficult proposition.
Here, we show how photoassociation in the 
vicinity of a Feshbach resonance enhance molecular formation rates 
by several orders of magnitude. We illustrate this effect in 
heteronuclear systems, and find giant rate coefficients even in 
deeply bound levels. We also give a simple analytical expression for 
the photoassociation rates, and discuss future applications of the
Feshbach Optimized Photoassociation, or FOPA, technique.

\end{abstract}

\pacs{32.80.Pj}
%\keywords{}

\maketitle

In recent years, several techniques, ranging from Stark
decelerators to buffer-gas cooling, have been developed
to obtain cold molecules \cite{EPJD-special}. 
Such molecules are interesting for a range 
of applications \cite{Jones2006} in metrology, high 
precision molecular spectroscopy, 
or quantum computing \cite{qc-mol}.
However, forming ultracold stable molecules in deeply bound 
levels remains a challenge: most approaches give temperatures
 still considered
{\it hot} (roughly 100 mK - 1 K).
To reach the ultracold regime 
%routinely experienced by atoms 
(below 1 mK), 
direct laser cooling of molecules is usually not effective
due to their rich and complex level structure \cite{Fioretti1998}.
Instead, it is possible to create %ultracold 
molecules starting from ultracold atoms, via photo-association
(PA) or ``magneto-association" (MA). While PA occurs when 
two colliding atoms absorb a photon to form a molecule 
\cite{Jones2006},
MA takes advantage of magnetically tuned Feshbach 
resosances \cite{ma-review}. 

Over the last decade, 
PA has been widely used to study long range
molecular interactions and 
to probe ultracold gases \cite{Jones2006},
and MA to realize molecular condensates \cite{mol-bec}
and investigate the BEC-BCS cross-over regime \cite{bec-bcs}.
However, both methods usually lead to molecules in highly excited 
states. According to the Franck-Condon principle, electronic 
transitions in PA occur at large interatomic distances, 
leading to molecules in high rovibrational levels that 
can either decay by spontaneous emission or collisional quenching. 
To stabilize the molecules in their ground potentials,
one could use two-photon schemes \cite{Juarros2004},
or excited molecular states with long-range wells
that increase the probability density at short 
range. This latter solution requires the existense of 
double-well molecular potentials \cite{Fioretti1998} and cannot
be easily generalized. In MA, molecules
are produced by sweeping the magnetic field through
a Feshbach resonance, which occur 
when the energy of a colliding pair of atoms matches 
that of a bound level associated to a closed channel. 
The molecules produced by MA are in the uppermost
states near dissociation \cite{ma-review} and thus relatively
extended and fragile.

\begin{figure}[htbp]
\epsfxsize=8.0cm\epsfclipon\epsfbox{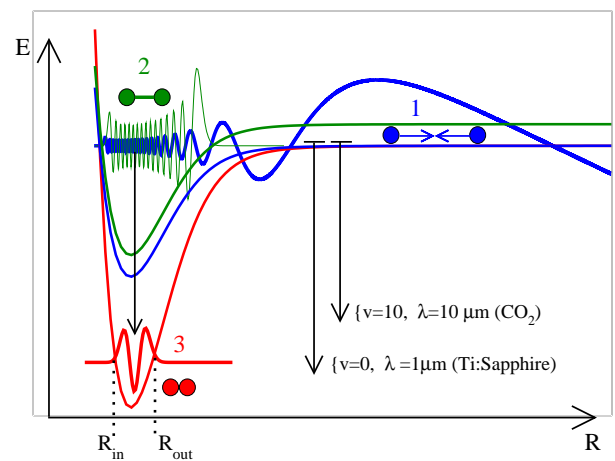}
\caption{FOPA: Colliding atoms (1) 
         interact via open (blue) and closed (green)
         channels due to hyperfine interactions. A 
         Feshbach resonance occurs when 
         a bound level (2) (green wave function) 
         coincides with the continuum 
         state (blue wave function). A photon (wavelength 
         $\lambda$) can associate the atoms into a 
         bound level $v$ (3) of the ground state potential 
         (red) with inner and outer classical turning points
         $R_{\rm in}$ and $R_{\rm out}$.}
\label{scheme}
\end{figure}

In this article, we investigate a new PA 
scheme which uses a magnetically induced Feshbach 
resonance \cite{ma-review} to enhance the probability 
density at short range. This Feshbach Optimized Photoassociation
(FOPA) allows transitions 
even to deeply bound levels (see Fig.~\ref{scheme}). 
%Mechanisms based on magnetically tuned 
Feshbach resonances 
and PA have been proposed to associate atoms \cite{vanAbeelen1999} 
%as well as convert an atomic BEC into a molecular BEC 
and convert an atomic into a molecular BEC
\cite{Kokkelmans2001}. However, as opposed to previous 
proposals \cite{pillet03-stwalley04}, FOPA takes advantage of the 
whole wave function in a full quantum coupled-channel calculation, 
and is thus more general than the Franck-Condon
principle.

Feshbach resonances are commonly found in both homonuclear 
and heteronuclear systems with hyperfine interactions.
We focus our attention on heteronuclear systems for which 
the presence of a permanent dipole moment allows transitions 
from the continuum directly to a rovibrational level $v$ of 
the ground electronic molecular states \cite{Juarros2006}
(see Fig.~\ref{scheme}). The corresponding photoassociation 
rate coefficient 
$K^v_{\rm PA}=\langle v_{\rm rel} \sigma^v_{PA}\rangle$ 
\cite{Jones2006,Juarros2004} depends on $v_{\rm rel}$, the relative 
velocity of the colliding pair, and on $\sigma^v_{PA}$, 
the PA cross section. The bracket stands for an average over 
the distribution of $v_{\rm rel}$, here a Maxwell-Boltzmann 
distribution characterized by the temperature $T$
\cite{note-distribution}. 
%If we assume
%a low laser intensity $I$, and ultracold temperatures where only 
%the $s$-wave contributes significantly, we can then express 
%the maximum rate coefficient as \cite{Juarros2006}	
At low laser intensity $I$ and ultracold temperatures, where only 
the $s$-wave contributes significantly, the maximum rate coefficient 
(neglecting saturation) is \cite{Juarros2006}	
\begin{equation}
   K^v_{\rm PA}=\frac{8\pi^3}{h^2}\frac{I}{c}
                \frac{e^{-1/2}}{Q_T}| \langle\phi_{v,J=1}
                  |D(R)|\Psi_{\epsilon,l=0}\rangle|^2 \; ,
  \label{rate}
\end{equation}
where $Q_T=(2\pi\mu k_B T/h^2)^{3/2}$, and $D(R)$ 
is the appropriate dipole moment for the transition
between the initial $|\Psi_{\epsilon,l=0}\rangle$
and final $|\phi_{v,J=1}\rangle$ states corresponding
to the $s$-wave $(l=0)$ continuum wave function of the colliding pair
and the populated bound level $(v,J=1)$ wave function.
Here $k_B$, $h$, and $c$, are the Boltzman
and Planck constants, and the speed of light in vacuum, respectively. 

We determine $|\Psi_{\epsilon,l=0}\rangle$ by solving the 
Hamiltonian for two colliding atoms in a 
magnetic field \cite{ma-review,Moerdijk1995} :
\begin{equation}
   H=\frac{p^2}{2\mu}+ V_C + \sum_{j=1}^{2}H^{\rm int}_j \;.
\label{Hamiltonian}
\end{equation}
Here, $V_C=V_0(R)P^0+V_1(R)P^1$ is the Coulomb interaction, 
decomposed into singlet ($V_0$) and triplet ($V_1$) molecular 
potentials, with the associated projection operator
$P^0$ and $P^1$. 
The internal energy of atom $j$, 
$  H^{\rm int}_j=\frac{a_{\rm hf}^{(j)}}{\hbar^2}\vec{s}_j\cdot\vec{i}_j + 
  (\gamma_e\vec{s}_j - \gamma_n\vec{i}_j)\cdot \vec{B}$,
consists of the hyperfine 
and Zeeman contributions, respectively.
%\begin{equation}
%\end{equation}
Here $\vec{s}_j$ and $\vec{i}_j$ are the electronic and nuclear 
spin of atom $j$ with hyperfine constant
$a_{\rm hf}^{(j)}$, and $\vec{B}$ is the magnetic field. 
Since the nuclear gyromagnetic factor $\gamma_n$ is three orders 
of magnitude smaller than $\gamma_e$, we neglect it in our 
calculations. 

%\vspace{0.5cm}
\begin{figure}[htbp]
 \epsfxsize=8.0cm\epsfclipon\epsfbox{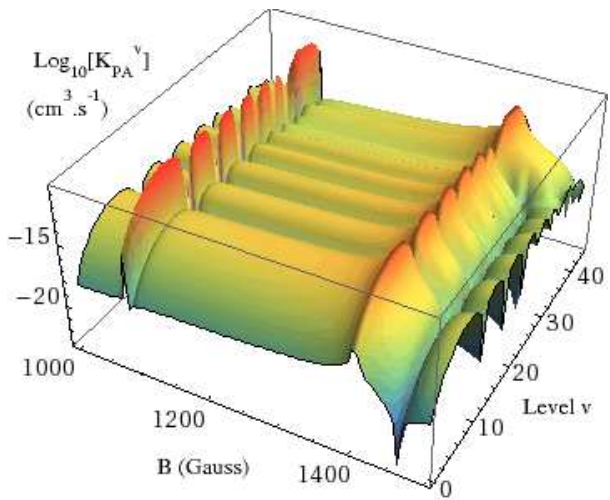}
\caption{$K^v_{\rm PA}$ in cm$^3$/s vs. the $B$-field 
         ($T=$ 50 $\mu$K, $I=1$ W/cm$^2$)
         for various levels $(v,J=1)$ of the LiNa X$^1\Sigma^+$
         potential, starting from 
         $^6$Li$(f=\frac{1}{2},m=-\frac{1}{2})$ and 
         $^{23}$Na$(f=1,m=-1)$. Two Feshbach resonances at 1081 and 
         1403 Gauss enhance the PA rates by several orders of 
         magnitude.}
\label{fig1}
\end{figure}

%\begin{figure}[htbp]
%\epsfxsize=8.0cm\epsfclipon\epsfbox{LiNa2x3_contour.eps}
%\caption{2D projection of Fig.~\ref{fig1}. The maximum 
%         $K^v_{\rm PA}$ does not vary much with the target 
%         level $v$, as opposed to the minimum values.}
%\label{fig2}
%\end{figure}

We solve for $|\Psi_{\epsilon,l=0}\rangle$ by using the Mapped 
Fourier Grid method \cite{Kokoouline1999}, and by
expanding it
onto the basis constructed from the hyperfine states of both atoms, 
\begin{equation}
\label{wf1}
 |\Psi_{\epsilon,l=0}\rangle=\sum_{\alpha=1}^{N} \psi_{\alpha}(R)
 \lbrace | f_1,m_{1}\rangle\otimes|f_2,m_{2}\rangle\rbrace_{\alpha} \; ,
\end{equation}
where $\vec{f}_j=\vec{i}_j+\vec{s}_j$ is the total spin 
of atom $j$, and $m_j$ its projection on the magnetic 
axis. Here, $\psi_{\alpha}(R)$ stands for the radial wave function associated 
with channel $\alpha$ labeled by the quantum numbers $f_i$, $m_{i}$; 
the Hamiltonian (\ref{Hamiltonian}) couples channels 
with the same total projection $M=m_{1}+m_{2}$. 
As an example, we consider forming LiNa in the 
ground X$^1\Sigma^+$ electronic state from atoms 
initially in the $^6$Li$(f=\frac{1}{2},m=-\frac{1}{2})$ 
and $^{23}$Na$(f=1,m=-1)$ states ($\alpha =1$), using the 
potentials of Ref.\cite{lina-feshbach}. Eight 
channels with total $M=-\frac{3}{2}$ are coupled, and we found
two Feshbach resonances at 
1081 and 1403 Gauss. Fig.~\ref{fig1} displays $K^v_{\rm PA}$ 
as a function of the $B$-field into different levels $(v,J=1)$
at $T=$ 50 $\mu$K and $I=1$ W/cm$^2$.
%at a temperature 
%of 5 $\mu$K and laser intensity of 1 W/cm$^2$. 
Near a resonance,  $K^v_{\rm PA}$ is drastically enhanced by 
up to five orders of 
magnitude, even for the lowest $(v < 10)$ levels. For typical 
densities $(n_{\rm Li}=n_{\rm Na}\sim 10^{12}$ cm$^{-3}$) 
and an illuminated volume $V$ of 1 mm$^3$,
$N_v=n_{\rm Li}n_{\rm Na}VK^v_{\rm PA}=2\times 10^6$ molecules are 
are formed in $v=0$ at 1403 Gauss
(neglecting back-stimulation \cite{Juarros2004}).

\begin{figure}[htbp]
\epsfxsize=8.0cm\epsfclipon\epsfbox{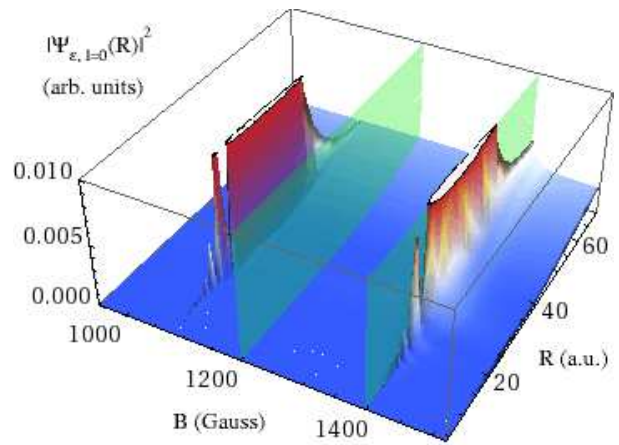}
\caption{Probability density $|\Psi_{\epsilon,l=0}(R)|^2$ vs.
         $B$. As $B$ nears a resonance, $|\Psi_{\epsilon,l=0}|^2$ increases 
         sharply (truncated above 0.01). Examples of $|\Psi_{\epsilon,l=0}|^2$ off 
         and on resonance (green planes at 1200 and 1400 Gauss, respectively)
         are shown in Fig.~\ref{FOPA}.}
\label{wf}
\end{figure}  

\begin{figure}[htbp]
\epsfxsize=8.0cm\epsfclipon\epsfbox{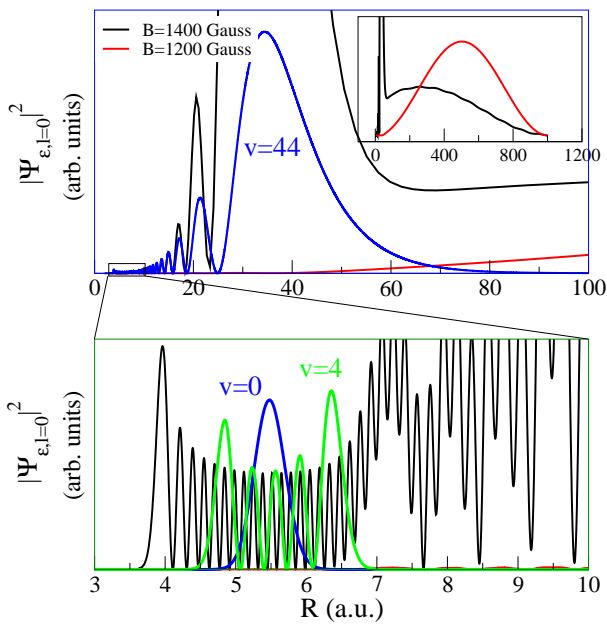}
\caption{Probability density on- (black) and off- (red) resonance. The top
         panel shows that $|\Psi_{\epsilon,l=0}|^2$ gets a peak
         near $R\sim 40$ a.u. on resonance (inset). The upper bound 
         level $v=44$ of the singlet ground state is also depicted.
         The off-resonance
         density is negligible for $R<50$ a.u. The bottom panel
         illustrates the inner region where the more deeply 
         bound target levels $v$ are located (e.g. $v=0$ and 4).
         Again, $|\Psi_{\epsilon,l=0}|^2$ is sizable on resonance 
         and negligible off-resonance. }
\label{FOPA}
\end{figure}

These giant formation rates can be understood by the sharp
increase in the amplitudes of the radial wave functions $\psi_{\alpha}$
in the vicinity of a Feshbach resonance. In Fig.~\ref{wf},
we show the total probability density $|\Psi_{\epsilon,l=0}(R)|^2$ 
as a function of $B$. As the magnetic field $B$
nears either of the resonances at 1081 and 1403 Gauss, 
$|\Psi_{\epsilon,l=0}(R)|^2$ increases by several order of magnitudes 
(qualitatively the same for all channels). 
%The nodal structure of this channel remains practically unchanged, 
%although this is not always the case. 
Fig.~\ref{FOPA} 
shows the total initial probability density 
$|\Psi_{\epsilon,l=0}(R)|^2$ on and off resonance 
($B=1400$ and 1200 Gauss, respectively):
%At large separation,
%$|\Psi_{\epsilon,l=0}(R)|^2$ is not much affected (see top panel,
%inset), the effect of the resonance becoming visible at shorter 
%distance, where a large peak appears near 40 $a_0$. 
%although it is modified at large separations, 
the main effect of the resonance is the
appearance of a large peak at shorter distance near
40 $a_0$ (see top panel, inset).
This peak is 
roughly located at the classical outer turning point $R_{\rm out}$ 
of the bound state associated to the closed channel, usually one 
of the uppermost bound levels. This is apparent in the top panel, 
where this peak almost coincides with the outer lobe of 
$|\phi_{v=44,J=1}(R)|^2$, the uppermost bound 
level of X$^1\Sigma^+$. 
We also observe that the off-resonance probability 
density is very much reduced when compared to on-resonance, 
leading to a very weak overlap integral in $K^v_{\rm PA}$. 
The lower panel shows the short separation range, where the on 
resonance probability density is much larger than the off resonance 
case, leading to a subtantial overlap integral in $K^v_{\rm PA}$ 
with deeply bound levels ({\it e.g.}, $v=0$ or 4). We also note 
the more complicated nodal structure of $|\Psi_{\epsilon,l=0}(R)|^2$, 
a direct result of the hyperfine mixing of the entrance channel 
$\alpha =1$ with all other channels ($\alpha = 2, \dots , 8$).

Analytical results are obtained with a two coupled channel model of reduced $\mu$, 
in which the wave function $\psi_1$ of the continuum state associated to 
the open channel 1 (with potential $V_1$) is coupled to the wave function 
$\psi_2$ associated 
to the closed channel 2 (with $V_2$) \cite{Friedrich} 
\begin{equation}
 -\frac{\hbar^2}{2\mu}\frac{d^2}{dR^2}{\psi_1 \choose \psi_2} 
 + \left(\begin{array}{cc} V_1 & V_{1,2} \\ V_{2,1} & V_{2} \end{array} \right)
   {\psi_1 \choose \psi_2}=E{\psi_1 \choose \psi_2} \;.
\label{eq:coupled}
\end{equation}
We assume both coupling terms $V_{1,2}$ and $V_{2,1}$ to be real, and fix
the threshold $E_1$ of channel 1 at $E=0$. If the couplings were switched 
off, the solution for the open channel 1 would be $\psi_1\rightarrow\psi_{\rm reg}$
while the closed channel 2 would have a bound state $\psi_2\rightarrow\psi_0$ 
with energy $E_0$. A resonance occurs when $E$
is near the energy $E_0$ of $\psi_0$. The analytical solutions 
for Eq.(\ref{eq:coupled}) are then \cite{Friedrich}
\begin{eqnarray}
   \psi_1 (R)\! & = & \! \psi_{\rm reg}(R) + \tan\delta\; \psi_{\rm irr}(R)\; , \nonumber \\
             \! & \stackrel{R\rightarrow \infty}{=} & \! \frac{1}{\cos\delta}\sqrt{\frac{2\mu}
         {\pi\hbar^2 k}}\sin(kR+\delta_{\rm bg}+ \delta) \;, \\
   \psi_2 (R)\! & = & -\sqrt{\frac{2}{\pi\Gamma}}\sin\delta\; \psi_0 (R)\;.
\end{eqnarray}
Here $\delta_{\rm bg}$ and $\delta$ are the background 
and resonant phase shifts, while $k=\sqrt{2\mu E}/\hbar$.
The asymptotic regular and irregular solutions
are
$\psi_{\rm reg}=\sqrt{\frac{2\mu}{\pi\hbar^2 k}}\sin(kR+\delta_{\rm bg})$
and $\psi_{\rm irr}=\sqrt{\frac{2\mu}{\pi\hbar^2 k}}\cos(kR+\delta_{\rm bg})$.
Finally, the width $\Gamma (E)$ of the resonance may vary slowly with $E$.

Here, scanning the $B$-field is equivalent to scanning $E$,
since the position $E_0$ of the bound state in channel 2 is shifted by
the Zeeman interaction. To first order in $k$, the $s$-wave phase shifts 
are related to the scattering length $a$ by 
$\tan(\delta + \delta_{\rm bg}) = -k a$, with $\delta_{\rm bg}= -k a_{bg}$
and  \cite{Moerdijk1995}
\begin{equation}
   a = a_{bg}\left( 1-\frac{\Delta}{B-B_0}\right) \;,
\label{eq:as}
\end{equation}
where $a_{\rm bg}$ is the background scattering 
length of the pair of atoms (which can slowly vary with $B$), 
$B_0$ is the position of resonance, and
$\Delta$ is related to  $\Gamma (E)$ \cite{Moerdijk1995}. 
Introducing the analytical solutions into 
Eq.(\ref{rate}) leads to
\begin{equation}
 K^v_{\rm PA}=K^v_{\rm off} \left| 1+ C_1 \tan\delta + C_2 \sin\delta \right|^2 \; ,
\label{var_rate}
\end{equation}
where $K^v_{\rm off}=\frac{8\pi^3}{h^2}\frac{I}{c}
\frac{e^{-1/2}}{Q_T}|\langle\psi_v|D|\psi_{\rm reg}\rangle|^2$ 
is the off-resonance rate coefficient  
($\delta = 0$) with $\psi_v$ the final (target) 
state,  
$C_1=\langle \psi_v |D| \psi_{\rm irr}\rangle / 
     \langle \psi_v |D| \psi_{\rm reg}\rangle$ relates 
to the open channel 1, while the coupling to the bound state
$\psi_0$ in the closed channel 2 is given by
$C_2=-\sqrt{2/\pi\Gamma}\langle \psi_v |D| \psi_0\rangle/ 
     \langle \psi_v |D| \psi_{\rm reg}\rangle$.

The relative importance of $C_1$ and $C_2$ depends on
the nodal structure of $\psi_v$, $\psi_{\rm reg}$,
$\psi_{\rm irr}$, and $\psi_0$. Unless $R_{\rm out}$ of 
$\psi_v$ accidentally coincides with a 
node in $\psi_{\rm reg}$ or $\psi_{\rm irr}$, the overlap 
integral of $\psi_v$ with both $\psi_{\rm reg}$ and 
$\psi_{\rm irr}$ are of the same order, leading to 
$|C_1|\sim 1$. The relative size of $C_2$ can be controlled by
the target level $v$. For a deeply bound level, 
$R_{\rm out}$ is
at short separation where the overlap with $\psi_{\rm reg}$ 
is small while the overlap with $\psi_0$ can be substantial
leading to $|C_2|\gg |C_1|$. For very extended levels $v$,  
$R_{\rm out}$ of $\psi_v$ is at large separation and 
the overlap with $\psi_0$ less important, leading to $|C_2|\ll |C_1|$. 
Naturally, these behaviours might differ for specific levels $v$. 

\begin{figure}[htbp]
\begin{center}

\epsfxsize=8.0cm\epsfclipon\epsfbox{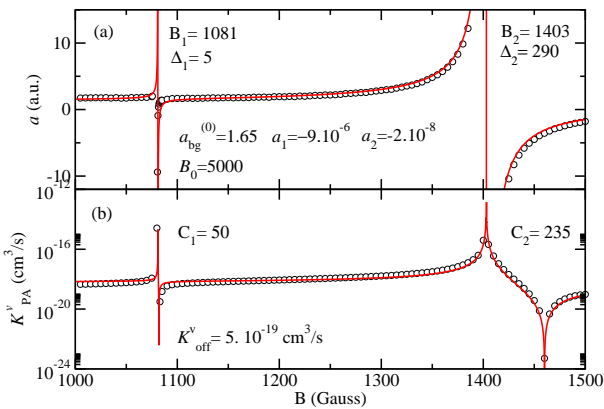}
\caption{In (a): scattering length $a$ for full coupled problem (circles)
         and the fit using $a=a_{\rm bg}(B)(1-\frac{\Delta_1}{B-B_1}-
         \frac{\Delta_2}{B-B_2})$ with $a_{\rm bg}(B)=a_{\rm bg}^{(0)}
         + a_1 (B+B_0) + a_2 (B+B_0)^2$. In (b): $K^v_{\rm PA}$
         for $v=0$ at 50 $\mu$K and 1 W/cm$^2$ (circles) and
         the simple
         formula (\ref{var_rate}) using 
         $\tan\delta=ka_{\rm bg}(B)(\frac{\Delta_1}{B-B_1}+
         \frac{\Delta_2}{B-B_2})$.
         The numerical parameters are given in each plot \cite{note-c2}.}
\label{fig2}
\end{center}
\end{figure}

The generalization of Eq.(\ref{var_rate}) to several coupled 
channels is straightforward as we simply add their contributions.
%from other closed and open channels. 
Furthermore, we find 
that only two or three channels contribute 
significantly to give these giant formation rate coefficients. 
In Fig.~\ref{fig2}, we show $K^v_{\rm PA}$ for the ground vibrational 
level $v=0$ with the same parameters used in 
Fig.~\ref{fig1}. The top panel depicts the scattering length 
$a$ with the two Feshbach resonances
and its analytical fit. The bottom panel
compares the exact numerical results using eight coupled channels 
with the simple expression (\ref{var_rate}).
In both cases, the agreement is impressive. We verified that similar
agreement was obtainable for other levels $v$, indicating the broad
and general validity of Eq.(\ref{var_rate}).

In conclusion, we showed that it is possible to use Feshbach Optimized
Photoassociation (FOPA) to form ``real" ultracold molecules, {\it i.e.} 
in deeply bound levels, in large quantities. In fact, the
rate coefficient increases by several orders of magnitude, leading
to giant formation rates near Feshbach resonances. We
applied this concept to LiNa, an heteronuclear system with a dipole
moment. In addition, we gave a simple analytical model describing the
FOPA technique. As opposed to other proposals using the Franck-Condon 
principle for transition near the turning point $R_{\rm out}$ of
the closed channel, FOPA takes advantage of the full 
wave function and its amplification in the vicinity of a Feshbach 
resonance, making it a general technique. In fact, FOPA could be 
used to do the spectroscopy of more deeply bound levels of excited 
electronic states which are usually not reachable by standard PA, 
bridging the gap between traditional spectroscopy for 
deep levels and PA of high-lying levels realized with ultracold 
atoms. Also, by targeting levels $v$ for which $C_1$ or $C_2$ is 
dominant, it is possible to determine the parameters of the scattering
length ($a_{\rm bg}$, $\Delta$, and $B_0$) by pure
spectrocopic measurements. This offers an accurate non-destructive
method to first detect a Feshbach resonance, and then obtain
the scattering length parameters. Finally, we note that this enhancement
will be present in other manifestation of Feshbach resonances, such
as those obtained via electric fields \cite{roman} or magnetic dipolar 
interactions ({\it e.g.}, in Cr \cite{cr}). This is a very general technique
which can be applied to bosonic, fermionic or mixed species, where 
Feshbach resonances exist. 

This research was supported by the U.S. Department of Energy,
Office of Basic Energy Sciences. 

%\newpage


\begin{thebibliography}{}
\bibitem{note-distribution}
  This $T$-dependence is valid at all $T$ for mixed species.
\end{thebibliography}
\end{document}